\documentclass[sigconf]{acmart}

\setcopyright{none}

\newcommand{\quotebox}[2]{\textit{``#2''}\ifthenelse{\equal{#1}{}}{}{ \mbox{-}~#1}}

\usepackage{color}
\usepackage{nameref}
\usepackage{multirow}
\usepackage{tabularx} 
\usepackage{censor}
\usepackage{nameref}

\usepackage[most]{tcolorbox}

\usepackage{xcolor}          

\definecolor{lightblue}{RGB}{0, 0, 100}

\newtcolorbox{MyBox}{
  colback=white,
  colframe=lightblue,
  fonttitle=\bfseries,
  coltitle=black,
  sharp corners,
  boxrule=1pt,
  left=5pt,
  right=5pt,
  top=5pt,
  bottom=5pt,
  breakable
}

\definecolor{purplish}{HTML}{D8DFE3}
\definecolor{purplishlight}{HTML}{EBEFF3}
\definecolor{purplishdark}{HTML}{FF7F50}
\definecolor{purplishdark}{HTML}{FF7F50}
\definecolor{myorange}{HTML}{FFA500}

\newtcolorbox[auto counter,number within=section]{rqbox}[2]{
    nameref=#1,
    title=\small{#1}, 
    enhanced,
    attach boxed title to top left={yshift=-6pt, xshift=8pt},
    boxed title style={size=small,boxsep=1pt},
    colframe=purplishdark,colback=white,colbacktitle=purplishdark,
    boxsep=2pt,left=2pt,right=2pt,top=6pt,bottom=2pt,middle=2pt
}

\newtcolorbox[auto counter,number within=section]{promptbox}[2]{
    nameref=#1,
    title=\small{#1}, 
    enhanced,
    attach boxed title to top left={yshift=-6pt, xshift=8pt},
    boxed title style={size=small,boxsep=1pt},
    colframe=myorange,             
    colback=white,                 
    colbacktitle=myorange,        
    boxsep=2pt,left=2pt,right=2pt,top=6pt,bottom=2pt,middle=2pt
}

\begin{document}

\title{Software Testing with Large Language Models: An Interview Study with Practitioners}

\author{Deolinda Santana}
\affiliation{%
  \institution{CESAR School}
  \city{Recife}
  \state{PE}
  \country{Brazil}
}
\email{mdos@cesar.school }

\author{Cleyton Magalhaes}
\affiliation{%
  \institution{UFRPE}
  \city{Recife}
  \state{PE}
  \country{Brazil}
}
\email{cleyton.vanut@ufrpe.br}

\author{Ronnie de Souza Santos}
\affiliation{%
  \institution{University of Calgary}
  \city{Calgary}
  \state{AB}
  \country{Canada}
}
\email{ronnie.desouzasantos@ucalgary.com}

\begin{abstract}
\textit{Background:} The use of large language models in software testing is growing fast as they support numerous tasks, from test case generation to automation, and documentation. However, their adoption often relies on informal experimentation rather than structured guidance. \textit{Aims:} This study investigates how software testing professionals use LLMs in practice to propose a preliminary, practitioner-informed guideline to support their integration into testing workflows. \textit{Method:} We conducted a qualitative study with 15 software testers from diverse roles and domains. Data were collected through semi-structured interviews and analyzed using grounded theory-based processes focused on thematic analysis. \textit{Results:} Testers described an iterative and reflective process that included defining testing objectives, applying prompt engineering strategies, refining prompts, evaluating outputs, and learning over time. They emphasized the need for human oversight and careful validation, especially due to known limitations of LLMs such as hallucinations and inconsistent reasoning. \textit{Conclusions:} LLM adoption in software testing is growing, but remains shaped by evolving practices and caution around risks. This study offers a starting point for structuring LLM use in testing contexts and invites future research to refine these practices across teams, tools, and tasks.
\end{abstract}



\maketitle

\section{Introduction}
\label{sec:introduction}

Large Language Models (LLMs) are AI systems trained on vast amounts of text and code, enabling them to generate human-like language and support a wide range of software development tasks \cite{hou2024large, fan2023large}. As a result, LLMs have seen growing adoption in software engineering, assisting with code generation, debugging, documentation, and maintenance activities \cite{michelutti2024systematic, zheng2025towards}. Initially used for small, isolated tasks, they are now being integrated more broadly into development workflows, often via IDE-integrated tools or conversational assistants such as ChatGPT and GitHub Copilot \cite{nam2024using, khojah2024beyond}. Currently, developers rely on LLMs not only to accelerate repetitive work but also to explore alternative solutions, understand unfamiliar code, and inform design decisions \cite{rasnayaka2024empirical, jahic2024state}. This shift reflects a growing recognition of LLMs as general-purpose assistants in professional software development.

In the context of software testing, LLMs are being used for a variety of tasks, including unit test generation, bug reproduction, test script writing, and code behavior explanation \cite{wang2023software, santos2024we, kang2023large, li2025evaluating, schafer2023empirical, fakhoury2024llm, mathews2024test}. They are particularly valuable in test automation and debugging, where they can rapidly produce test cases or clarify error messages \cite{santos2024we, wang2023software}. Additionally, testers with limited programming experience are using LLMs to bridge technical gaps, especially during test analysis and documentation \cite{santos2024we}. However, most existing studies focus on functional testing at the implementation level. Earlier stages, such as testing planning and design, and non-functional testing activities, such as performance and security validation, remain underexplored \cite{boukhlif2024llms, santos2024we}.

While LLMs offer clear benefits by reducing manual effort and increasing accessibility \cite{hou2024large, santos2024we, michelutti2024systematic}, their use also raises concerns. These include hallucinated outputs, vague reasoning, and a lack of verification mechanisms \cite{fan2023large, wang2023software, hou2024large}. Their nondeterministic behavior makes it difficult to fully trust their recommendations, posing challenges in structured testing workflows that depend on predictability and formal quality assurance methods \cite{labiche2000testing, umar2019comprehensive}. In this context, the use of LLMs in software testing is currently shaped by informal practices, community advice, and gray literature \cite{santos2024we}. There are no established frameworks or standards to guide their integration into the testing life cycle. To address this gap, our study investigates how LLMs are used in real-world testing environments and proposes a preliminary set of practice-based guidelines. These recommendations aim to support responsible integration of LLMs into industry workflows, aligning emerging LLMs capabilities with foundational testing principles.

\smallskip
{\narrower \noindent \textit{\textbf{RQ.} How do software testers use LLMs in their work, and what steps can be drawn from their experiences to guide integration into testing activities?} \par}
\smallskip

To answer this question, we conducted interviews with 15 software testing professionals, analyzing their practical experiences with LLMs in daily testing tasks. We focused on the challenges they encountered and the strategies they developed to work effectively with these tools. As a result, this paper contributes:
\begin{itemize}
\item An empirical report on how software testing professionals are using LLMs in industry settings.
\item A preliminary guideline for incorporating LLMs into software testing workflows, grounded in real-world experience.
\end{itemize}

From this introduction, this study is organized as follows. Section~\ref{sec:background} gathers recent studies about LLMs in software testing. Section~\ref{sec:method} describes our survey methodology. In Section~\ref{sec:findings}, we present our findings, which are discussed in Section~\ref{sec:discuss}. Section~\ref{sec:limitations} discussed threats to validity. Finally, Section~\ref{sec:conclusions} summarizes our contributions and final considerations.

\section{Testing with LLMs} \label{sec:background}
  
As the interest in applying LLMs to software engineering grows, a growing body of research has begun to investigate their applicability, benefits, and limitations in testing contexts \cite{boukhlif2024llms, yu2023llm}. The integration of LLMs into software testing has emerged as a promising direction for addressing long-standing challenges in test automation, test maintenance, and knowledge transfer across testing tasks. While traditional automated testing approaches are often rigid and require significant effort to adapt to changing systems, LLMs offer flexible, prompt-based interactions that can support a variety of tasks across different testing levels \cite{wang2023software, michelutti2024systematic}. 

LLMs are currently being used to assist in tasks such as unit test generation, bug reproduction, test script synthesis, test oracle creation, and test documentation \cite{schafer2023empirical, wang2025automating, kang2023large, mathews2024test, santos2024we}. Unit test generation is one of the most widely studied applications, where models are prompted with source code or method descriptions and return compilable test cases. Some approaches complement LLM-generated tests with post-processing or refinement mechanisms to improve correctness and coverage \cite{yuan2024evaluating, li2025evaluating}. In bug reproduction, LLMs are prompted with natural language bug reports and tasked with generating inputs that recreate the issue, showing potential to support maintenance and debugging workflows \cite{kang2023large}. Additionally, LLMs have been explored as tools for writing oracles, explaining test failures, and supporting exploratory testing in poorly documented systems \cite{wang2025automating, mathews2024test}.

Although LLMs can lower entry barriers and reduce effort, their integration into testing activities presents specific challenges. Generated outputs may be syntactically correct but semantically misaligned with the intended software behavior, or include hallucinated assertions and faulty logic \cite{hou2024large, fan2023large}. The lack of consistency, determinism, and verification mechanisms further challenges their use in formalized quality assurance processes. Some studies have proposed ranking and filtering techniques to address these shortcomings, while others have emphasized prompt engineering and iterative refinement to improve output reliability \cite{wang2025roadmap, yuan2024evaluating}. There is also evidence that the utility of LLMs is highly context-dependent, with performance varying across tasks, programming languages, and domains \cite{boukhlif2024llms}.

In this sense, professionals in software testing are beginning to adopt LLMs in informal and exploratory ways, often guided by community-shared prompts or trial-and-error experimentation \cite{santos2024we}. Many use LLMs as collaborative assistants rather than standalone tools, helping with ideation, code explanation, and documentation rather than fully automating test tasks. This trend is especially visible among testers with limited programming backgrounds, who benefit from LLMs’ ability to bridge technical knowledge gaps and improve their confidence in dealing with complex code \cite{santos2024we, fakhoury2024llm}. However, in the absence of formal guidelines, their adoption remains ad hoc, raising concerns about overreliance, trust, and integration into established testing workflows \cite{li2025evaluating}.

\section{Method}
\label{sec:method}

This study investigates how software testing professionals integrate large language models (LLMs) into their workflows, with the goal of identifying key practices that inform a preliminary guideline for LLM adoption. We used a qualitative interview design~\cite{ralph2020empirical} composed of three phases: (1) sampling and recruitment to ensure a range of relevant testing professionals; (2) data collection through semi-structured, one-on-one interviews; and (3) a multi-stage coding process using grounded theory principles and thematic analysis~\cite{charmaz2014constructing, cruzes2011recommended}. This structure allowed us to explore how testers interpret and integrate LLMs into daily activities, navigate emerging challenges, and learn through hands-on use.

\subsection{Sampling and Participant Recruitment}

We employed a combination of convenience and snowball sampling~\cite{baltes2022sampling} to reach participants with experience in software testing and LLM usage. Initial participants were drawn from the researchers’ professional networks and testing communities. Early interviewees were then invited to refer colleagues with diverse roles, levels of seniority, and LLM exposure. As analysis progressed, we used theoretical sampling to seek participants who could expand or contrast emerging patterns~\cite{charmaz2014constructing}. We contacted over 45 professionals and successfully recruited 15 individuals. Many declined due to non-disclosure agreements (NDAs) of their companies or internal restrictions on LLM usage. Our final sample included manual testers, test automation engineers, QA analysts, and test managers from multiple contexts and industries. We collected demographic and role information at the start of each interview to characterize the participant pool and ensure coverage of a range of professional profiles.

\subsection{Data Collection and Instrumentation}
All interviews were conducted synchronously, one-on-one, and online between October 2024 and January 2025. Interviews lasted between 25 and 45 minutes. Conversations were audio-recorded with permission and later transcribed verbatim. We used a semi-structured interview guide that combined open-ended questions with follow-ups to explore emergent responses. The guide was developed iteratively based on prior work on LLMs in software testing and refined through a pilot with two professionals and feedback from an expert in qualitative methods. The guide focused on usage contexts, tasks supported by LLMs, perceived benefits and limitations, and interactions with organizational practices. Table~\ref{tab:interview} presents the final interview script. No compensation was provided, and no incentives were offered to avoid participation bias.

\begin{table}[ht]
  \caption{Interview Guide}
  \label{tab:interview}
  \vspace{-8px}
  \begin{tabularx}{\linewidth}{p{1.8cm} X}
    \toprule
    Section & Questions \\
    \midrule

    General &
    1. Can you describe your experience with the use of LLMs in software testing? \\
    \midrule

    LLM Usage in Testing & 
    2. What models or tools have you used for testing purposes?\\
    & 3. For which types of testing activities have you applied LLMs? \\
    \midrule

    Benefits and Challenges & 
    4. What are the main benefits you have observed from using LLMs in testing? \\
    & 5. What challenges or concerns have you encountered when using LLMs for testing? \\
    \midrule

    Strategies and Best Practices & 
    6. What strategies or best practices have you adopted to maximize the effectiveness of LLMs in testing? \\
    & 7. What measures can be taken to mitigate the risks and challenges associated with using LLMs in software testing? \\
    \midrule

    Impact on Quality & 
    8. How do you assess the impact of LLMs on software quality? \\
    \midrule

    Demographics & 
    9. How do you identify in terms of gender? \\
    & 10. What is your highest level of education? \\
    & 11. How long have you worked in software testing? \\
    & 12. What is your level of seniority in software testing (junior, mid-level, senior)? \\
    & 13. In which country do you reside, and where is your company headquartered? \\
    & 14. Are your activities conducted remotely, on-site, or in a hybrid format? \\
    & 15. Do you hold any certifications related to software testing or quality assurance? \\
    & 16. Can you share details on projects you have worked on, including technologies, methodologies, project scope, or your specific role? \\
    
    \bottomrule
  \end{tabularx}
  \vspace{-10px}
\end{table}

\subsection{Data Analysis}

We analyzed the data using a grounded theory–informed thematic analysis~\cite{charmaz2014constructing, cruzes2011recommended}, incorporating open, focused, and theoretical coding. All transcripts were reviewed line-by-line by the first author using open coding to identify discrete events, practices, and judgments related to LLM usage. Codes were labeled close to the data (e.g., \textit{“Human-in-the-loop Validation”}). In the focused coding phase, similar open codes were grouped into broader categories such as \textit{Evaluating the Output}. Theoretical coding was then used to relate these categories and organize them into a five-step structure that reflected how participants described integrating LLMs into their testing workflows. We maintained a codebook with noted coding decisions and discussed ambiguous cases among the authors for consistency. Direct quotations are used throughout the paper to support credibility and transparency, illustrating themes while preserving participants’ voices. We followed reflexive practices and peer debriefing to mitigate interpretive bias.

\subsection{Saturation and Rationale for Sample Size}
We assessed theoretical saturation by tracking the emergence of new concepts throughout the interview process. By the twelfth interview, we observed that no substantially new categories were emerging, and subsequent interviews reinforced existing patterns. This decision was supported by the repeated presence of the same five steps across participants and saturation of variation within each step. Given the richness of the data and the constraints of participant availability due to LLM-related policies, we concluded the study with 15 interviews. This sample size aligns with prior qualitative software engineering studies and satisfies the requirements for saturation in focused exploratory research~\cite{guest2006howmany}.

\subsection{Ethical Considerations}

This study was reviewed and approved by the Research Ethics Board at the third author's institution. Participants were provided with information about the study’s goals, procedures, and confidentiality measures before consenting to participate. No identifiable data were collected, and participant anonymity was maintained throughout. We do not release the raw dataset due to confidentiality concerns, but anonymized quotations are included within the findings to support transparency and traceability. Researchers remained attentive to power dynamics, and participants had the option to skip questions or withdraw at any time.

\section{Results} \label{sec:findings}
The participants in this study belong to a diverse cohort of software testers, representing different levels of experience, educational background, industries, and work arrangements. The group consists of 15 professionals, with 27\% (4/15) classified as junior testers, 33\% (5/15) as mid-level, and 40\% (6/15) as senior testers. In terms of education, 40\% (6/15) hold a bachelor’s degree, 47\% (7/15) have completed postgraduate studies (master’s or specialization), and 13\% (2/15) are still pursuing an undergraduate degree. The participants work across multiple industries, ensuring a broad representation of real-world software testing environments. The most common sectors include finance, healthcare, e-commerce, AI-driven applications, and embedded systems. Their work arrangements are also diverse, with 40\% (6/15) working fully remotely, 20\% (3/15) in hybrid settings, and 40\% (6/15) in on-site roles. In terms of professional certifications, several participants (60\%) hold professional credentials, including CTFL, CTAL, and Scrum certifications, while others rely on practical experience and specialized courses. Their expertise spans manual and automated testing, utilizing industry-standard tools such as Selenium, Cypress, Robot Framework, and Appium. Additionally, at work, many participants engage in AI-assisted testing, incorporating LLMs like ChatGPT, Gemini, Copilot, and Blackbox AI into their workflows. This diversity in experience, industry, and testing methodologies provides a well-rounded perspective on the use of LLMs in software testing. Table \ref{tab:Demographics} summarizes our participants' backgrounds.

\begin{table}[ht]
  \caption{Participant Demographics (N=15)}
  \label{tab:Demographics}
  \vspace{-8px}
  \begin{tabularx}{\linewidth}{p{1.8cm} X}
    \toprule

    Experience Level & 
    Junior Tester: 4 individuals \newline
    Mid-level Tester: 5 individuals \newline
    Senior Tester: 6 individuals \\
    \midrule

    Education & 
    Bachelor’s Degree: 6 individuals \newline
    Postgraduate Studies: 7 individuals \newline
    Currently in Undergraduate: 2 individuals \\
    \midrule

    Work \newline Arrangement & 
    Fully Remote: 6 individuals \newline
    Hybrid: 3 individuals \newline
    On-site: 6 individuals \\
    \midrule

    Certifications & 
    Holds Professional Certifications: 9 individuals \newline
    No Formal Certifications: 6 individuals \\

    \bottomrule
  \end{tabularx}
  \vspace{-10px}
\end{table}

\subsection{LLMs in Testing: Common Uses and Adoption Trends}
Our findings indicate that LLMs are primarily used for test case design, automation, and learning, where AI tools help in the execution of repetitive tasks and improve efficiency. The use of LLMs in documentation, execution, and validation is also growing, but applications such as test metrics, log analysis, and test evidence evaluation remain less common. Based on the experience of the testers, we observed that while LLMs have been reducing manual workloads, their integration into more advanced testing activities is still in its early stages. More specifically, testers reported using LLMs in the following areas:

\begin{itemize}
    \item \textbf{Test case and scenario creation (8/15):} LLMs assist in generating structured test cases based on requirements, user stories, or system specifications. This helps to improve workflows and ensure broad test coverage without the need for relying on fully manual scenario writing. Regarding this use, P6 explains: \textit{''I basically use it to create test cases, they were simple test cases, so I worked with a refined prompt.''} Similarly, P13 highlights: \textit{''I really like using it for test case creation (...) I usually copy parts of the story, where there is no sensitive data.''}  

    \item \textbf{Test automation (7/15):} LLMs are used to generate, refine, and optimize automated test scripts, assisting with debugging and improving script efficiency in tools such as Selenium and Cypress. These capabilities enable testers to accelerate test execution and improve coverage. P2 describes how they leverage LLMs for test automation: \textit{''So, we use Copilot and ChatGPT to make some improvements (...) there was a script written in Robot Framework, and we are converting it to C\#.''}  

    \item \textbf{Research and learning (7/15):} LLMs assist testers in quickly retrieving information, troubleshooting issues, and staying updated on best practices and emerging methodologies. Instead of manually searching forums or documentation, testers use LLMs to get immediate, contextualized answers. On this usage, P1 states: \textit{''When you had a problem, you used to go to online forums to find a solution (...) with ChatGPT, I just input my problem, and it immediately gives me direction.''} P12 adds: \textit{''I use ChatGPT for personal learning, sometimes to create a study plan when I don’t even know where to start.''}  

    \item \textbf{Test documentation (5/15):} LLMs help testers draft, structure, and standardize test documentation, including test templates, reports, and execution plans. This reduces manual effort and ensures consistency across teams. For example, P11 discussed: \textit{''Even for documentation automation, it is possible to develop systems using AI to automate the creation of documentation.''}  

    \item \textbf{Test execution (5/15):} Some testers integrate LLMs into automated test execution workflows, where LLMs assist in running pre-defined tests and analyzing results. This enables more efficient execution of repetitive test cases while identifying potential failures. In this context P5 commented: \textit{''Today, we use the GPT-4o LLM to assist in the creation of test cases, as we follow the IEEE 829 standard''.} 

    \item \textbf{Generating acceptance criteria (4/15):} LLMs assist in defining clear and structured acceptance criteria, helping teams formalize validation requirements based on system specifications and user stories. P8 highlighted: \textit{''Scenario creation, based on rules, and acceptance criteria also based on rules.''}  

    \item \textbf{Test evidence analysis (3/15):} LLMs are used to process and compare test evidence, including UI consistency checks and defect detection in graphical interfaces and mobile applications. This helps testers evaluate differences in test results and detect visual inconsistencies more efficiently. P5 described: \textit{''I have also used LLMs: Gemini-1.5-flash and GPT-4o to analyze images taken from test evidence in order to extract and compare results.''}  
\end{itemize}

In addition to these primary uses, LLMs were also mentioned in less frequent activities. Some testers reported using AI for \textbf{requirement generation (2/15)}, \textbf{bug description and validation (2/15)}, \textbf{test case validation (2/15)}, and \textbf{metric evaluation for test cases (2/15)}. A few also mentioned \textbf{log analysis (1/15)} and \textbf{code conversion (1/15)}, where AI assists in detecting system anomalies and adapting test scripts for different programming languages. While these uses exist, they appear to be less common among testers and may still be in early adoption stages.

\subsection{How LLMs Enhance Software Testing: Reported Advantages}

Our findings show that practitioners gain various benefits from incorporating LLMs into software testing. The most frequently reported advantages include improved time efficiency, faster problem resolution, enhanced productivity, and greater test coverage. Additionally, LLMs assist in reducing effort in test creation, minimizing errors, and supporting learning and training. These benefits contribute to more effective workflows, a reduction in repetitive tasks, and overall improvements in testing quality. Below, we described the key benefits cited by the participants:

\begin{itemize}
\item \textbf{Time Efficiency (9/15):} Testers frequently mentioned that LLMs help save time when searching for solutions, creating test cases, and performing testing tasks. Rather than manually navigating through online resources or consulting colleagues, they receive instant guidance. LLMs also help structure test cases and optimize scripts, reducing redundant work. P1 explains: \textit{''I think LLMs help a lot with learning because they save time. Instead of searching through forums, I just put my problem into ChatGPT, and it gives me direction right away.''} Similarly, P10 states: \textit{''If you're starting a new project and aren’t familiar with it, using ChatGPT in this context saves a lot of time and optimizes the work.''}

\item \textbf{Enhanced Productivity (4/15):} Several testers noted that LLMs made them more productive at work by automating many tasks and improving their test design and execution. By providing structured suggestions, LLMs reduce the need for testers to manually refine scenarios and acceptance criteria. P8 explains: \textit{''It’s mainly about productivity. You don’t waste time thinking too much—LLMs give you something, and if it makes sense, you use it. If not, you discard it and move on.''}  

\item \textbf{Improved Test Coverage (3/15):} LLMs were reported to help testers expand their test coverage by generating diverse scenarios and increasing validation points. Testers reported that AI-assisted case generation enables more thorough testing while maintaining efficiency. P15 highlights: \textit{''The biggest gain is agility in creating test cases, validations, and expanding automated test coverage. We can test more possibilities.''}  

\item \textbf{Learning and Training Support (3/15):} LLMs assist in knowledge acquisition by providing structured guidance for understanding new concepts, tools, and methodologies. Some testers also use LLMs to facilitate onboarding, helping new team members work independently. P15 explains: \textit{''We trained interns who were starting in the technology field, but we didn’t have to closely monitor them. They could handle tasks that normally would have taken them months to learn.''}  
\end{itemize}  

Less frequently mentioned benefits include \textbf{faster problem resolution (2/15)}, where testers found that LLMs assist in troubleshooting by providing direct answers, reducing the time spent debugging or searching through documentation. \textbf{Reduced effort in test creation (2/15)} was also noted, as LLMs help structure test cases based on requirements and user stories, minimizing the manual effort required to define test scenarios. Finally, \textbf{error reduction (1/15)} was cited by one tester, who observed that LLMs contribute to more reliable testing workflows by automating validation steps, ultimately minimizing mistakes in the testing process.

\subsection{Navigating the Challenges of LLM Adoption in Software Testing}

While testers identified several benefits of integrating LLMs into testing workflows, they also encountered context-specific challenges related to software quality assurance. The most recurrent issues include accuracy and hallucination in generated artifacts, confidentiality risks in test data, limited prompting proficiency, over-reliance on automated reasoning, and organizational resistance to tool integration. Each of these challenges affects distinct stages of the testing process.

\begin{itemize}
\item \textbf{Accuracy and Hallucination in Test Artifacts (9/15):} Testers frequently observed that LLMs produced incorrect, incomplete, or logically inconsistent test cases, test data, or defect explanations. These inaccuracies risk propagating false positives or negatives in automated test suites. P4 remarked: \textit{“Sometimes it gives a wrong answer, then it hallucinates, like, it starts doing things I didn’t ask for.”} Several participants indicated that unreliable LLM outputs required manual validation and regression testing to ensure correctness.

\item \textbf{Data Privacy and Security in Test Environments (7/15):} The use of project specifications or production data to generate test inputs created apprehension about exposing confidential information. Testers stressed that prompts often include details from internal systems, increasing the risk of data leakage to external APIs. P3 noted: \textit{“The main concern, especially in the workplace, is information leaks.”} Practitioners emphasized the need for sanitizing inputs and deploying models on private infrastructure when feasible.

\item \textbf{Prompting and Tool Adoption Difficulties (3/15):} Adapting LLMs for testing tasks, such as generating boundary cases or summarizing logs, required learning how to construct precise prompts. Several testers reported a lack of established prompting patterns for testing-specific needs like equivalence partitioning or coverage analysis. P3 explained: \textit{“The main challenge at first was learning how to use the tool. It felt a bit strange.”} Similarly, P14 added: \textit{“The challenge I had in the beginning was getting accurate answers (with my prompts).”}

\item \textbf{Risk of Cognitive Over-Reliance (3/15):} Some participants feared that excessive dependence on LLM suggestions might erode testers’ analytical judgment, especially in exploratory and risk-based testing. They worried that automated reasoning could replace systematic test design skills or critical assessment of failure conditions. P12 expressed: \textit{“My biggest concern (…) is becoming dependent on AI, like, not being able to create a simple scenario without it.”}

\item \textbf{Organizational and Process Barriers (1/15):} One tester highlighted the challenge of aligning LLM-assisted testing with existing quality assurance policies and obtaining management approval for tool integration. P15 explained: \textit{“I think the challenge was showing the company that it added value so they would invest in it.”} Such resistance often stems from uncertainty about validation standards and return on investment in testing automation.

\end{itemize}

Despite these difficulties, testers described several mitigation strategies. To address hallucinations, they triangulate outputs across multiple tools or use retrieval-augmented generation to ground responses in project documentation. To preserve confidentiality, they anonymize test data or operate LLMs on restricted servers. Many practitioners invest time in prompt engineering to refine test generation and automate low-risk cases while maintaining human review for critical tests. Integration efforts are also coordinated with governance and security teams to ensure compliance with organizational standards. Overall, testers recommend partial automation—using LLMs to support, rather than replace, human reasoning—to sustain accuracy, accountability, and trust in the testing process.

\subsection{A Preliminary Practitioner-Informed Guideline to LLM Use in Software Testing}

\begin{figure*}[!ht]
    \centering
    \includegraphics[width=17cm]{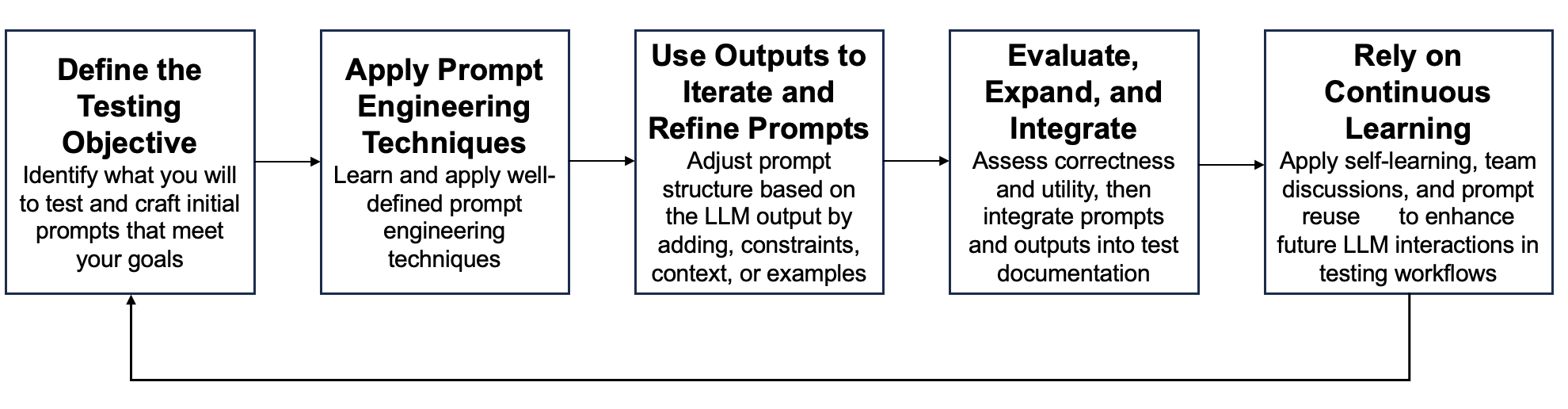}
    \caption{LLM into Software Testing}
    \label{fig:guide}
\end{figure*}

To understand how software testers incorporate LLMs into their daily testing activities, we examined the usage patterns and lessons described by participants and organized the reported practices into a logical sequence. This structure is not a fixed process but a set of interconnected actions that reflect how practitioners employ and refine LLM-assisted testing. The resulting guideline describes empirically grounded steps, illustrated by quotations in Table~\ref{tab:quotes} and Figure~\ref{fig:guide}.

\begin{enumerate}
\item \textbf{Define the Testing Objective and Structure the Prompt:}
Testers began by identifying the testing objective, such as generating test cases for a specific feature, analyzing logs, or refactoring existing scripts. They then structured prompts to express that objective clearly and specify the expected testing artifact. This initial alignment between the test goal and the prompt ensured that the model produced outputs relevant to the testing task. P01 and P03 described framing the testing problem explicitly so that the model would generate focused and actionable outputs.

\item \textbf{Apply Prompt Engineering Techniques for Testing Tasks:}
After defining the testing goal, testers refined their prompts using structured techniques to guide the LLM’s behavior. They mentioned prompt engineering strategies such as the CARE method, and one-shot or few-shot examples to provide the model with reference test cases. These techniques reduced ambiguity and improved the precision of generated artifacts such as boundary-value tests or refactored code snippets. P04 and P05 emphasized that developing these strategies required study and experimentation similar to learning a new testing tool.

\item \textbf{Iterate and Refine Testing Prompts:}
Participants described a cycle of prompt refinement similar to regression or exploratory testing. When the LLM produced incomplete or misaligned test cases, they incrementally adjusted prompts by adding context, constraints, or expected output formats. This iterative interaction resembled test design refinement, where successive runs led to prompts that generated more consistent and reusable test cases. P06 and P11 characterized this process as shaping the output until it matched their testing expectations.

\item \textbf{Evaluate, Expand, and Integrate LLM Outputs into the Testing Pipeline:}
Once an acceptable output was produced, testers evaluated its accuracy, readability, and maintainability before integrating it into their testing assets. They treated LLM-generated material such as refactored automation code or draft test cases as preliminary artifacts to be reviewed and extended. P08 and P06 explained how they validated the generated code, compared it against their own implementation, and incorporated useful portions into Cypress scripts or manual test documentation. This stage represents the transition from experimentation to controlled adoption within the testing workflow.

\item \textbf{Engage in Continuous Learning and Knowledge Sharing:}
Participants emphasized that effective use of LLMs for testing requires continuous learning. They regularly researched new prompting strategies, attended internal sessions on AI-assisted testing, and experimented with applying LLMs to increasingly complex testing scenarios. This ongoing learning enabled testers to align model usage with evolving testing demands and organizational quality goals. P12, P05, and P07 described these efforts as essential for improving testing efficiency without compromising analytical rigor.

\end{enumerate}

These steps portray how testers integrate LLMs into quality assurance practices through iterative, reflective, and human-supervised activities. Rather than treating LLMs as autonomous testing tools, practitioners use them to augment test design, code review, and automation tasks while maintaining human oversight to ensure correctness, completeness, and contextual relevance.

\begin{table}[ht]
  \caption{Participant Quotations}
  \label{tab:quotes}
  \vspace{-8px}
  \begin{tabularx}{\linewidth}{p{2.8cm} X}
    \toprule
    Step & Quotation \\
    \midrule

    Define the Testing Objective and Structure the Prompt Clearly &
    \textbf{P01}: \textit{“So yes, I try to be very objective. For example, I frame it like a problem, a development, and a conclusion. I present the problem so it can give me a clear solution.”} \\
    & \textbf{P03}: \textit{“I try to send a question that really expresses what I am testing, and that way I can get better use out of the tool.”} \\
    \midrule

    Apply Prompt Engineering Techniques &
    \textbf{P05}: \textit{“Improving prompts is something I have been dedicating time to—studying which methods I can apply to get the result I want. One method I have used is CARE.”} \\
    & \textbf{P04}: \textit{“I took a course on prompt engineering, and I also research prompt strategies online. There are some techniques, like one-shot, few-shot, and others I cannot recall now.”} \\
    \midrule

    Iterate and Refine Testing Prompts &
    \textbf{P06}: \textit{“When I was creating the prompt, for example, I saw that the first descriptions of the features did not generate a test case that matched what I expected, so I kept refining it, adding more information.”} \\
    & \textbf{P11}: \textit{“It starts getting shaped—and by the next time, maybe you ask again and it already gives you the structure you picked, the one that worked best for you.”} \\
    \midrule

    Evaluate, Expand the Output, and Integrate the LLM into the Pipeline &
    \textbf{P08}: \textit{“I give it a Cypress code, for example, and ask it to refactor it. I see if the refactoring makes sense—if it does, I accept it; if not, I stick with mine.”} \\
    & \textbf{P06}: \textit{“That was when I realized—it actually generates something I would have done. I would improve it later, sure, but it gave me a solid starting point, a skeleton of the test case I could adjust.”} \\
    \midrule

    Learn and Improve LLM Usage &
    \textbf{P12}: \textit{“The company I work at has initiatives—there is a release every day, and they host live sessions explaining how we can use Flow more effectively.”} \\
    & \textbf{P05}: \textit{“I have been researching a lot about AI technologies and trying to bring that into our testing work.”} \\
    & \textbf{P07}: \textit{“I use an iterative approach—starting with simple tasks and expanding to more complex ones as the LLM adapts to the project’s context.”} \\
    
    \bottomrule
  \end{tabularx}
  \vspace{-10px}
\end{table}

\subsection{Answering the Research Question}

Our findings answer our research question by showing that software testing professionals use LLMs through an iterative and reflective process integrated into tasks such as test creation, automation, documentation, and analysis. Aware of LLM limitations such as hallucinations and privacy risks, testers rely on human oversight to validate output. From these practices, we derived a five-step preliminary guideline for LLM adoption and usage: 1) define the testing objective and structure the prompt; 2) apply prompt engineering techniques; 3) iterate and refine testing prompts; 4) evaluate, expand, and integrate LLMs to the testing pipeline; and 5) continuous learning to improve LLM usage. These steps offer a practical foundation for more deliberate LLM use in testing workflows.

\section{Discussions} 
\label{sec:discuss}
While much of the current literature emphasizes what LLMs can achieve in testing, such as test generation, bug reproduction, or test script creation~\cite{hou2024large, michelutti2024systematic, boukhlif2024llms, kang2023large, yuan2024evaluating}, our study shifts attention to how testers engage with these tools in everyday settings. Participants described using LLMs to support various tasks, including interpreting error messages, drafting test evidence, and handling ambiguous scenarios, often as part of informal and adaptive routines. Rather than acting as replacements, LLMs were positioned as complementary aids, offering guidance and efficiency boosts without supplanting professional judgment. This perspective aligns with empirical findings showing LLMs are being incorporated to augment rather than automate testing activities~\cite{fan2023large, schafer2023empirical, mathews2024test}.

However, our study shifts the discussion from the potential and limitations of LLMs to the practical behaviors that testers adopt to make these tools useful to their work. Existing research tends to describe usage patterns, benefits, and risks at a high level, often focusing on controlled benchmarks ~\cite{michelutti2024systematic, wang2023software, li2025evaluating}. In contrast, our participants discussed how they iteratively worked with LLMs in real software projects by adjusting prompts, cross-verifying outputs, and incorporating model responses. This type of study has received comparatively little attention in the literature, even though they are important to understand responsible and effective adoption~\cite{fakhoury2024llm, santos2024we}. For instance, our participants described developing adapted practices and fallback strategies to deal with limitations such as hallucinations, poor domain alignment, or privacy constrains, behaviors that illustrate how LLM use is far from plug-and-play in professional testing environments.

Furthermore,the novelty of this study based on the articulation of these behaviors into a structured yet flexible set of five steps that represent how testers interact with LLMs in practice. While previous work has highlighted the need for LLM usage guidelines in testing~\cite{santos2024we, michelutti2024systematic, wang2025roadmap}, our contribution emerges directly from the field, grounded in the lived experience of professionals in real-world environments. These steps reflect an evolving practice, from the initial identification of a testing task suited for LLM use, to iterative refinement of prompts, evaluation of responses, contextual adaptation, and integration into broader workflows. Rather than prescriptive instructions, this preliminary guideline captures adaptable usage patterns that support more reflective and informed integration of LLMs across testing contexts.

Ultimately, this study centers in the human side of LLM adoption. Rather than asking what LLMs can do or how well they perform in abstract terms, we focused on how testers make sense of these tools, incorporate them into their work, and evaluate their outcomes in light of professional standards and industry demands. Our participants demonstrated an understanding of both the possibilities and limitations of LLMs, grounded not in technical evaluation but in everyday experience. This people-focused lens adds an important dimension to ongoing discussions in the field, suggesting that successful integration of LLMs into testing depends as much on practitioner expertise as on model capabilities.

\subsection{Implications for Research}
This study contributes to the growing body of empirical research on LLMs in software engineering by documenting how testing professionals integrate these tools into their workflows. Our study does not focus solely on what LLMs are technically capable of, but emphasizes how testers adapt and evaluate LLM outputs in real-world contexts. By exploring practitioner behavior, this work narrows the gap between academic studies and industry adoption. The five-step guideline we propose serves as an initial conceptual structure that reflects current practices, offering researchers a concrete starting point for theorizing LLM-supported testing from a socio-technical perspective.

Future research can build on these findings in multiple directions. Each step of the guideline opens opportunities to investigate how LLMs are applied across specific testing activities, such as exploratory testing, non-functional testing, or test evidence generation, and how these practices vary across domains, experience levels, or organizational settings. Moreover, the observed reliance on informal experimentation and prompt refinement highlights the need for educational strategies and tool support tailored to testers. Designing interventions that promote responsible and effective LLM usage and studying their impact in longitudinal or collaborative testing environments, represents a promising direction to further integrate LLMs into the evolving software testing landscape.

\subsection{Implications for Practice}
For industry practice, this study offers testing professionals and software organizations a practitioner-informed foundation for integrating LLMs into real-world testing workflows. The five-step guideline reflects common behaviors adopted by testers, such as refining prompts, reviewing outputs, and selectively incorporating model suggestions. Rather than prescribing rigid procedures, it captures adaptable strategies that can help teams assess and evolve their use of LLMs. This makes it especially valuable for organizations in the early stages of LLM adoption, offering them a way to reflect on current practices, identify areas for improvement, and reduce the uncertainty that often accompanies the introduction of emerging technologies.

In practical terms, the findings suggest concrete actions companies can take. These include facilitating structured training sessions focused on LLM prompting techniques, encouraging the reuse and documentation of successful prompts across teams, and promoting open discussion about the limitations and risks associated with LLM usage. Additionally, the guideline can assist practitioners in evaluating existing tools based on how well they support the behaviors testers find most useful. By supporting reflective and informed adoption, this work contributes to bridging the gap between exploratory LLM experimentation and sustainable practices in software testing.

\section{Threats to Validity}
\label{sec:limitations}
While we presented significant findings regarding how professionals engage with LLMs in software testing, it is important to acknowledge that this is a qualitative study, and due to its nature and the limitations of this type of method, including the number of participants interviewed, we do not claim generalization. Instead, we aimed for theoretical transferability, which means that our research allow others to consider, adapt, or implement these findings, considering the particularities of their own contexts.

Due to confidentiality concerns, particularly as some participants work in environments where LLM use is discouraged, we cannot share full interview transcripts. However, to support transparency, we included anonymized quotations to illustrate reported practices while preserving participant anonymity. Additionally, we followed a qualitative analysis process involving iterative coding and constant comparison to ensure that our findings, in particular the steps of the preliminary guideline, are grounded in the real-world data.

Finally, data collection concluded at the point of theoretical saturation, when new interviews reinforced existing categories rather than introducing additional ones. This approach strengthens the consistency of the reported patterns and is in line with the qualitative research guidelines \cite{ralph2020empirical}.

\section{Conclusions} 
\label{sec:conclusions}
Our study explored how software testing professionals integrate LLMs into their daily workflows, based on the experience of 15 testing professionals. Our findings highlight that testers do not treat LLMs as fully autonomous tools, but rather described an iterative process to incorporate them into software testing. This process begins with clearly defining the testing objective and structuring prompts to align with that goal. Testing professionals then apply prompt engineering strategies to improve response quality. When initial outputs fall short, testers refine prompts until the model delivers a more suitable result. Outputs are then evaluated and, when appropriate, adapted and integrated into testing artifacts such as automation scripts, documentation, or reports. Finally, testers emphasized continuous learning as an essential step to improve usage over time.

Given the documented limitations of LLMs in software engineering, such as hallucinations, vague reasoning, and lack of context retention, which were also concerns raised by testers in our study, having well-defined and transparent usage practices becomes important. The preliminary guideline we propose responds to this need by gathering practice-based activities that reflect how LLMs are currently being incorporated in real-world testing contexts. This guideline is not intended to be prescriptive but rather offers a grounded starting point for professionals navigating LLM adoption in testing. By encouraging reflection on current practices, raising awareness of potential risks, and supporting more intentional use, our study provides a practical foundation for integrating LLMs into testing workflows.



\balance

\bibliographystyle{ACM-Reference-Format}
\bibliography{bib}

\end{document}